\documentclass[a4paper]{jpconf}
\usepackage{graphics}
\usepackage{graphicx}
\begin{document}
\newcommand{\arh}{\ensuremath{ArH^+}}
\newcommand{\htwo}{\ensuremath{H_2}}
\newcommand{\htwoplus}{\ensuremath{H_2^+}}
\newcommand{\hhh}{\ensuremath{H_3^+}}
\newcommand{\otp}{ortho-to-para}

\newcommand{\thirdwidth}{0.3\textwidth}     
\newcommand{\miniwidth}{0.48\textwidth}     

\newcommand{\midiwidth}{0.6\textwidth}     
\newcommand{\sidewidth}{0.35\textwidth}     

\newcommand{\maxiwidth}{0.9\textwidth}     

\title{Spectroscopy and dissociative recombination of the lowest rotational states of H$_3^+$}

\author{A~Petrignani$^1$, H~Kreckel$^2$, M~H~Berg$^1$, S~Altevogt$^1$, D~Bing$^1$, H~Buhr$^3$, M~Froese$^1$,
J~Hoffmann$^1$, B~Jordon-Thaden$^1$, C~Krantz$^1$, M~B~Mendes$^1$, O~Novotn\'{y}$^1$,
S~Novotny$^1$, D~A~Orlov$^1$, S~Reinhardt$^{1,}$\footnote[4]{Present address: Max-Planck-Institut
f\"{u}r Quantenoptik, Garching, Germany}, and A~Wolf$^1$}

\address{$^1$ Max-Planck-Institut f\"{u}r Kernphysik, Saupfercheckweg 1, D-69117 Heidelberg, Germany}

\address{$^2$ Columbia University, 550 West 120th Street, New York, NY 10027, USA}

\address{$^3$ Department of Particle Physics, Weizmann Institute of Science, 76100 Rehovot, Israel}

\ead{Annemieke.Petrignani@mpi-hd.mpg.de}

\begin{abstract}
The dissociative recombination of the lowest rotational states of \hhh\ has been investigated at the storage ring TSR using a cryogenic 22-pole 
radiofrequency ion trap as injector. The \hhh\ was cooled with buffer gas at $\sim$15\,K to the lowest rotational levels, ($J,G$)=(1,0) and (1,1), 
which belong to the ortho and para proton-spin symmetry, respectively. The rate coefficients and dissociation dynamics of \hhh($J,G$) populations 
produced with normal- and para-\htwo\ were measured and compared to the rate and dynamics of a hot \hhh\ beam from a Penning source. The production of 
cold \hhh\ rotational populations was separately studied by rovibrational laser spectroscopy using chemical probing with argon around 55\,K. First 
results indicate a $\sim$20$\%$ relative increase of the para contribution when using para-\htwo\ as parent gas. The \hhh\ rate coefficient observed 
for the para-\htwo\ source gas, however, is quite similar to the \hhh\ rate for the normal-\htwo\ source gas. The recombination dynamics 
confirm that for both source gases, only small populations of rotationally excited levels are present. The distribution of 3-body 
fragmentation geometries displays a broad part of various triangular shapes with an enhancement of $\sim$12$\%$ for events with symmetric near-linear 
configurations. No large dependences on internal state or collision energy are found.
\end{abstract}

\section{Introduction \label{sec:intro}}
The \hhh\ ion is of utmost theoretical and astrophysical importance. It is the simplest polyatomic ion and thereby constitutes an important benchmark 
ion for the theory of both molecular structure and molecular collision dynamics. It is also a key ion in the universe as it initiates a network of 
reactions leading to the formation of complex molecules. The dissociative recombination (DR) of \hhh\ plays a crucial role as destruction mechanism 
for the \hhh\ ions \cite{mccall:2005,oka:2006} and as production mechanism for the neutral hydrogen fragments. The DR reaction has been extensively 
researched over the past decades \cite{larsson:2000,mccall:2006} and with the knowledge on the reaction increasing, the focus is moving to more and 
more detailed investigations. Internal-state dependences of the DR reaction, at zero as well as at elevated electron energies, are now being studied 
\cite{wolf:2006,greene:2006}. Disagreements on the magnitude of the DR rate coefficient of \hhh\ have been dominating the scene for quite some time 
\cite{oka:2000a}. Presently, the differences that have been observed are attributed to different internal-state distributions of the ions and consent 
between storage-ring experiments has been reached \cite{mccall:2004,kreckel:2005a}. Furthermore, previous discrepencies with afterglow studies seem to 
have been resolved recently and the latest theoretical calculations show good overall agreement with experiment 
\cite{glosik:2007,kokoouline:2003b,santos:2007}. However, the predicted rate coefficient still shows local discrepancies up to an order 
of magnitude around several electron energy regions. Again, differing internal-state distributions are given as possible explanation. In the cold 
interstellar space, the difference of the DR rate between the two lowest rotational levels, ($J,G$)=(1,0) and (1,1) at an energy difference 
corresponding to kT=33\,K only, may be of particular importance. Since for \hhh\ these two levels are equivalent with the ortho and the para symmetry 
of the total nuclear spin, respectively, we shortly denote the two levels as ortho- and para-\hhh. Previous experiments indicate that the DR rate 
coefficient of para-\hhh\ could be larger than that of ortho-\hhh\ for specifically the low electron energies prevailing in the cold interstellar 
medium \cite{kreckel:2005a}. In these experiments, para-\htwo\ was used as precursor gas and although the rotational populations of the \hhh\ ion beam 
could not be measured, an enhancement of the para-\hhh\ population is expected in this case. The higher para-\hhh\ rate was until recently in 
contradiction to theory, which predicted the opposite \cite{kokoouline:2003b,kokoouline:2005}. However, new theoretical results now state a roughly 
ten times higher rate for para-\hhh\ at low temperature \cite{santos:2007}.

The DR of cold \hhh\ has two fragmentation channels for low-energy electrons, a 3-body breakup containing only ground-state atoms and a 2-body 
breakup that includes the \htwo\ product molecule,
\begin{eqnarray*}
	\hhh(J,G) + e^{-}(E_d) & & \rightarrow H(1s) + H(1s) + H(1s)\,,		\\
								& & \rightarrow \htwo(\upsilon,J) + H(1s)	
\end{eqnarray*}

\noindent Here, the cold initial \hhh\ ion is given to be in a low rotational state of its ground vibrational state with $J$ the total angular 
momentum associated with the motion of the nuclei and $G=|K-l|$, where $K$ is the projection of $J$ onto the molecular axis and $l$ is the vibrational 
angular momentum. The product atoms are in their ground state, whereas the \htwo\ molecule may be rovibrationally, $(\upsilon,J)$, excited. Further 
dissociation channels become possible when additional energy, $E_d$, is introduced via the electrons. At typical molecular-cloud temperatures of 
10-60\,K, \hhh\ populates the two to three lowest rotational levels, ($J,G$)=(1,0), (1,1), and (2,2), of the vibrational ground state, where the (2,2) 
level belongs to the para spin symmetry.

We present a two-fold investigation of cold \hhh. The first focus is on the production and characterisation of rotationally cold \hhh\ populations 
using normal-\htwo\ and para-\htwo\ as parent gas. In order to probe the different rotational distributions, rovibrational spectroscopy on the \hhh\ 
ions in a 22-pole radiofrequency ion trap is implemented. The second focus is on the DR of \hhh($J,G$) populations with altered \otp\ ratio. 
High-resolution rate coefficients up to 0.4 eV are given and additionally compared to a hot \hhh\ beam to investigate rotational heating. The issue of 
internal excitation is addressed by fragment imaging. For the first time, the dissociation of \hhh\ were studied using the $3D$ imaging technique
\cite{strasser:2000}, and 3-body breakup from both $2D$ and $3D$ imaging are presented.

\section{Experimental Setup \label{sec:exp}}

\begin{figure}[h]
\includegraphics[width=\maxiwidth]{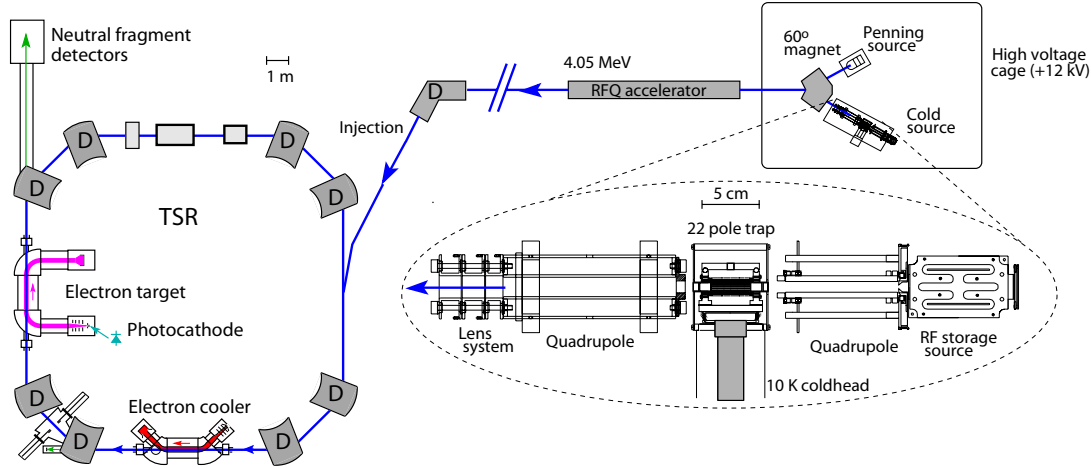}%
\caption{\label{fig:trap}The 22-pole trap setup used as injector for the ion storage ring TSR. The
\hhh\ ions are prepared in the trap, accelerated to 4.05 MeV, and injected into the ring. A cold
electron beam, produced with a liquid-nitrogen cooled photocathode \cite{orlov:2004}, is merged with the ion beam in the electron
target. The neutral fragments are detected either with a surface barrier counting detector or a
position-sensitive imaging system \cite{strasser:2002a,nevo:2007}.}
\end{figure}

The cold \hhh\ samples are prepared in a 22-pole radiofrequency (rf) ion trap setup, as shown in the inset of Figure \ref{fig:trap}. The ions are 
produced in a rf storage ion source \cite{teloy:1974} through electron-impact ionisation of \htwo\ and subsequent collisions of the \htwo\ with the 
hence produced \htwoplus. The \hhh\ ions are then guided through a rf quadrupole to the 22-pole rf ion trap \cite{gerlich:1995}.
Here, the ions are trapped radially by the 22 poles and axially by an entrance and exit electrode. The number of stored ions may be varied from a few 
to several millions of ions, whereas the storage time may be varied from ms to several seconds. During storage, the ions are buffer-gas cooled with 
helium at a chosen ambient temperature of 10-60\,K. For more details on the 22-pole trap setup, the reader is referred to Refs. 
\cite{mikosch:2004,kreckel:2005b}. In the present work, the setup is used as a stand-alone device for probing the \hhh($J,G$) populations implementing 
rovibrational spectroscopy and it is used as an ion injector for the storage ring TSR to the study of the DR of the cold \hhh\ (see Figure 
\ref{fig:trap}). Unfortunately, the conditions required for the spectroscopy measurements are different from those of the DR measurements. As such, 
care has to be taken when deploying information from the laser-probed populations for the TSR measurements.

\subsection{Spectroscopy Experiments \label{sec:trap}}
To verify the buffer-gas cooling inside the 22-pole trap and to probe the rotational-state populations, an action spectroscopy 
scheme is used. To this purpose, rovibrational spectroscopy in the infrared is implemented as \hhh\ lacks any stable electronically excited states. As 
a sensitive probe of the vibrational transitions, a laser-induced chemical reaction with argon as reactant is used. The charge exchange reaction,
\begin{equation}
    \hhh(\upsilon=0) + Ar \rightarrow \arh + \htwo - 0.57 \,\,\textrm{eV} \label{eq:Ar}
\end{equation}

\noindent is endothermic by 0.57 eV for ground-state \hhh\ ions. However, if at least two vibrational quanta are excited in \hhh, the reaction 
becomes energetically possible. Once the ions in the trap are cooled down to the chosen temperature, a tunable infrared diode laser of $\sim$1400 nm 
is used to trigger transitions from the lowest rotational states of \hhh\ to the third harmonic of the bending-mode vibration. The laser-excited 
ions will react with the abundant argon in the trap to form \arh, which is trapped as well and used as spectroscopic signal. The ions are extracted 
from the trap, analysed by a mass spectrometer, and directed to a scintillation detection system capable of counting single ions with near-unity 
efficiency. These spectroscopy measurements are performed at a minimum temperature of 55\,K to keep the argon in the gas phase. For more details, see 
Ref. \cite{mikosch:2004}. 

The action spectroscopy scheme has been used before to probe cold \hhh\ produced with normal-\htwo\ \cite{mikosch:2004}. Here, also the \hhh\ 
rotational distribution produced with para-\htwo\ is presented. Moreover, important improvements have 
currently been implemented to enhance the cooling of the \hhh\ in the trap as well as to enhance the \arh\ spectroscopy signal. 
Gas purities, laser power (increased by a factor of 10), and frequency resolution have now been refined. As a result, the typical \arh\ count at 
resonance is 1.5-2 per trap filling of 300 \hhh\ ions, an order of magnitude higher than before. The number densities of the helium and argon gas 
inside the trap have been optimised to $\sim$10$^{14}$ cm$^{-3}$ and $\sim$10$^{12}$ cm$^{-3}$, respectively. The hydrogen number density, 
$\sim$10$^{10}$ cm$^{-3}$, and the trap storage time, 250 ms, were unchanged.

\subsection{Dissociative Recombination Experiments \label{sec:tsr}}
The DR measurements were performed with the 22-pole trap as injector of the cold ion beam for the storage ring TSR.
The trap temperature was set to 15\,K in order to simplify the DR measurements, as only the lowest para and ortho rotational states, (1,0) and 
(1,1), are populated at that temperature. The number of \hhh\ ions had to be hugely increased to the maximum trap filling of around $2\cdot10^6$ in 
order to supply a sufficient DR count rate \cite{kreckel:2003}. The \hhh\ \otp\ ratio is dominantly determined by collisions with \htwo, therefore the 
\htwo\ number density (\mbox{$10^{10}$ cm$^{-3}$}) was kept the same as in the spectroscopy experiments. The helium number density was increased to 
\mbox{$\sim$10$^{15}$ cm$^{-3}$} to compensate for the slightly reduced trap storage time of 100 ms. After this cooling time, the ions were released 
to be accelerated and stored in the TSR at an ion-beam energy of 4.05 MeV. The transmission of the ion beam was about 10\%, resulting in very low 
counting rates (in a later storage-ring experiment the transmission could be improved to 50\%). The ring storage time was set to 10 s during which the 
ion beam was merged in the electron target with an ultracold electron beam produced by a liquid-nitrogen cooled photocathode \cite{orlov:2004}. The 
transversal and longitudinal temperatures of this electron beam were T$_{\perp}=1$ meV and T$_{\parallel}=30$ $\mu$eV, respectively. The electron 
density was roughly $9\cdot10^5$ cm$^{-3}$. Rate-coefficient curves were measured with a new large surface barrier detector of 10x10 cm$^2$ at 
collision energies between 0 and 0.4 eV by varying the electron detuning energy, $E_d$. The dynamics of the DR reactions were investigated at $E_d=0$ 
and 6 meV through the $2D$ \cite{strasser:2002a} and, as a first for \hhh, the $3D$ \cite{strasser:2000,nevo:2007} imaging technique. The electron 
cooler was set to 0 eV collision energy to continuously cool the ion beam and, for the first 0.5 s of storage, the electron target was additionally 
set to 0 eV in order to speed up the initial translational phase-space cooling of the beam. This initial cooling time was deliberately kept low in 
order to minimise possible internal heating-effects from the toroidal regions, where the electron beam is no longer parallel to the ion beam and 
elevated detuning energies may cause rotational excitation of the ions. Only for the 0-eV imaging measurements, the electron cooler was switched off 
after this time, whilst the target remained at 0 eV.

\section{Results \label{sec:results}}

\subsection{Rovibrational Spectroscopy \label{sec:spec}}

\begin{figure}[h]
\includegraphics[width=\midiwidth]{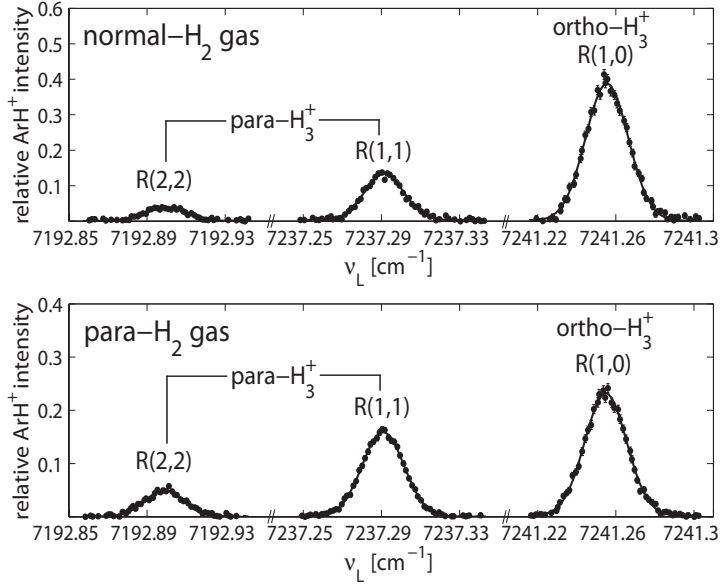}\hspace{1.5pc}%
\begin{minipage}[b]{\sidewidth}
\caption{\label{fig:spec} The measured absorption profiles for the three observed transitions in the $(0,3^1)\leftarrow(0,0^0)$ vibrational overtone 
band of \hhh\ produced with normal-\htwo\ (upper panel) and para-\htwo\ (lower panel). The signal is derived by averaging over many trap fillings.}
\end{minipage}
\end{figure}

Figure \ref{fig:spec} shows the observed \arh\ signal for the transitions from the three lowest rotational states present at 55\,K to the third 
vibrational level. The upper figure results from \hhh\ ions produced with normal-\htwo, the lower from \hhh\ ions produced with para-\htwo. We are 
currently still studying possible effects of the change of the precursor gas on the absolute normalisation of the signal. The relative normalised 
intensities for a given normal-\htwo\ or para-\htwo\ measurement, however, are reproducible and well under control. The intensities are normalised to 
the laser power of $\sim$12 mW, to a trap filling of 300 ions, and to an \arh\ lifetime of 25 ms, such that they may be coarsely compared to the 
previously published spectroscopy measurement \cite{mikosch:2004}.

The experimental changes mentioned before have now lead to improved signal strengths giving rise to nicely observable Gaussian profiles. The cooling 
of the ions also proved to be more efficient at the optimised helium pressure and much lower temperatures are observed than in 2004 
\cite{mikosch:2004}. The profile widths of the respective transitions denoted by $R(J,G)$ -- where $R$ stands for the ($J$+1) branch and $J,G$ stand 
for the rotational quantum numbers of the lower state -- are determined by Doppler broadening, which conforms a temperature of $\sim$67\,K for all 
transitions. The line-intensity ratios are determined by the given laser intensity, the Einstein coefficients, and the initial-level populations. 
Assuming the theoretical Einstein coefficients of 8.9, 4.9, and 8.2 s$^{-1}$ for $R(1,0)$, $R(1,1)$, and $R(2,2)$, respectively \cite{dinelli:1992}, 
the populations and hence \otp\ and para-to-para ratios can be extracted. The ``para rotational temperature'' is derived from the ratio 
between the two para transitions, $R(1,1)$ and $R(2,2)$, and matches the translational temperature of $\sim$67\,K for both ion-source gases. Thermal 
equilibrium seems to have been reached in both cases. The ``\otp\ temperature'' is derived from the ratio between the ortho transition, $R(1,0)$, and 
the para transitions, $R(1,1)$+$R(2,2)$. A significant decrease of the ortho-\hhh\ line relative to the para lines is visible when using para-\htwo\ 
as parent gas. The \otp\ temperature correspondingly decreases, namely from $\sim$140\,K to $\sim$36\,K. This clearly shows that under certain 
circumstances the \otp\ ratio can become non-thermal and can deviate significantly from the translational and rotational temperature of an ion sample. 
The population for the normal-\htwo\ parent gas conforms an \otp\ ratio of 60:40 using the above mentioned Einstein coefficients, whereas the \hhh\ 
sample produced with para-\htwo\ shows the opposite, an \otp\ ratio of 40:60. It seems that the use of para-\htwo\ as parent gas leads to enrichment 
of the rotational states belonging to the para-\hhh\ symmetry.

\subsection{Dissociative Recombination Rate Coefficients \label{sec:rate}}

\begin{figure}[h]
\begin{minipage}[t]{\miniwidth}
\includegraphics[width=\textwidth,height=14pc]{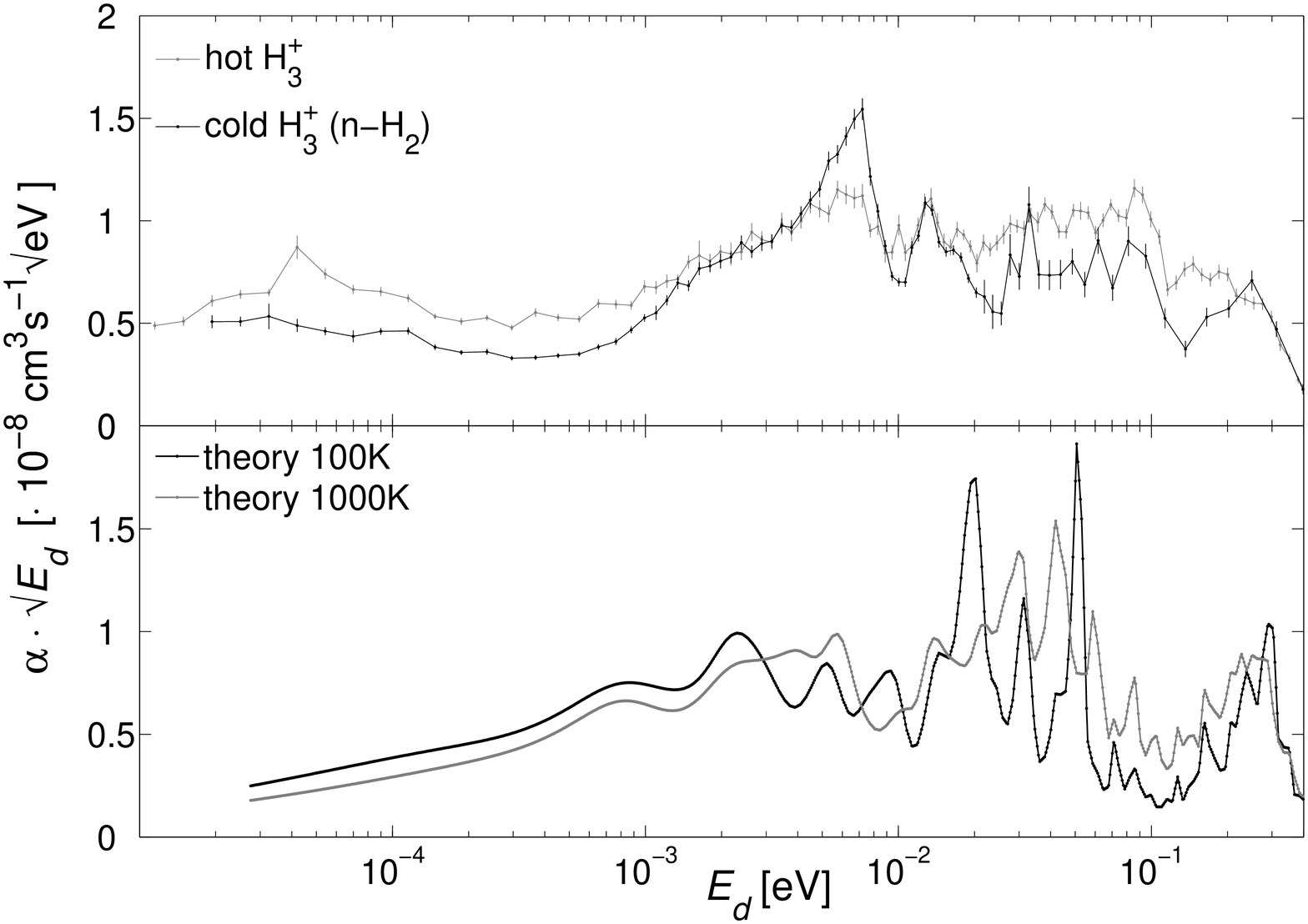}
\caption{\label{fig:Trates}The upper frame shows the measured reduced rate coefficients of the cold (black) and hot (grey) \hhh\ produced with 
normal-\htwo\ in the 22-pole trap and the Penning source, respectively. The lower frame shows the reduced theoretical rates for 100\,K (black) and 
1000\,K (grey), folded with the experimental electron temperature from Ref. \cite{santos:2007}.}
\end{minipage}\hspace{1.5pc}%
\begin{minipage}[t]{\miniwidth}
\includegraphics[width=\textwidth,height=14pc]{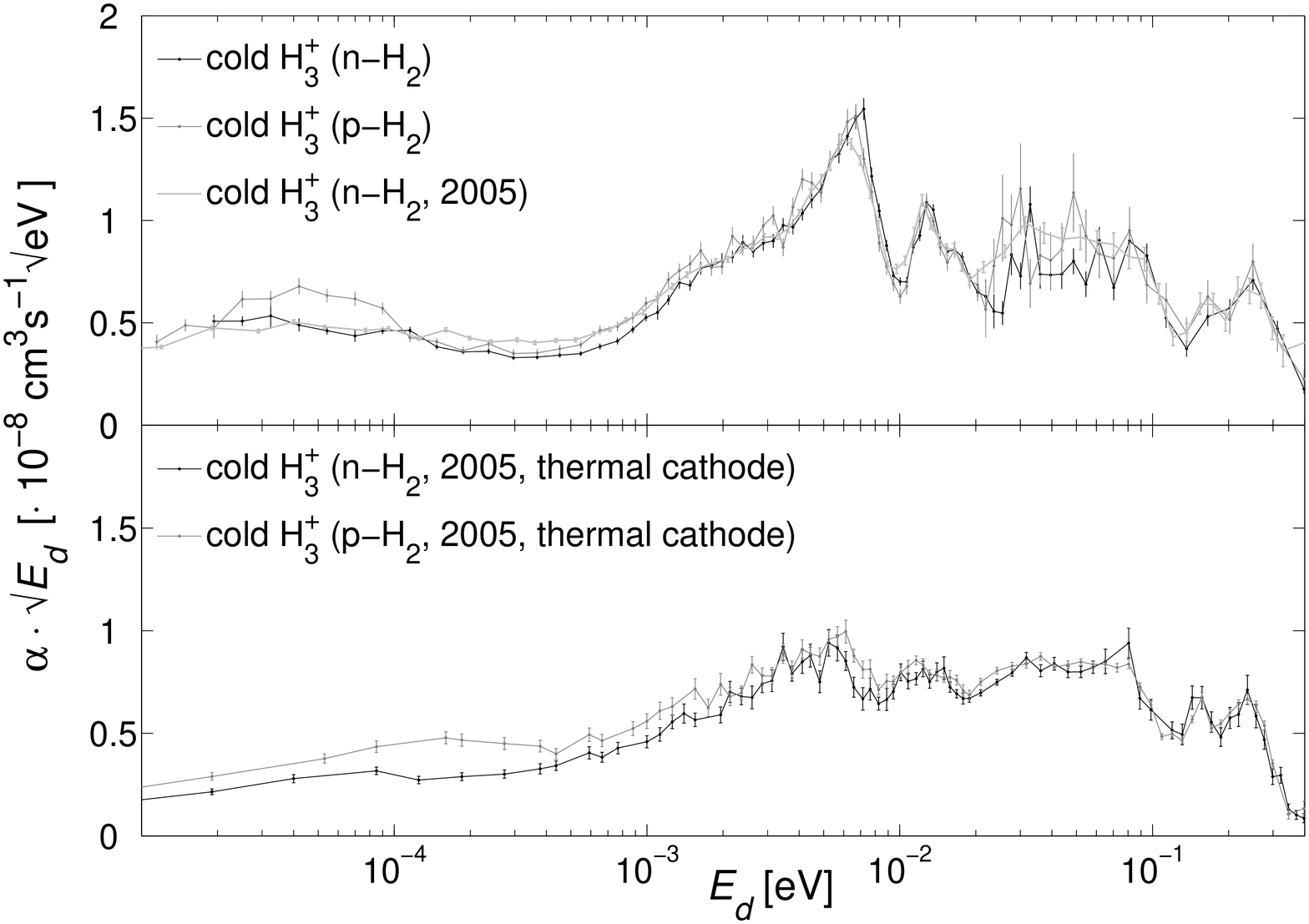}
\caption{\label{fig:rates}The upper frame shows the high-resolution reduced rates of \hhh\ produced with para-\htwo, present data (grey), and 
normal-\htwo, present data (black) and previous data (light grey) \cite{kreckel:2005a}. The lower frame shows the previously measured 
lower-resolution reduced rates of \hhh\ produced with para-\htwo\ (grey) and normal-\htwo\ (black) \cite{kreckel:2005a}.}
\end{minipage}
\end{figure}

As mentioned before, the 22-pole trap was used as an injector for the storage ring TSR, where the DR rate of the cold \hhh\ populations was 
investigated between 0 and 0.4 eV collision energy. Also, to investigate rotational heating, the DR rate of a hot \hhh\ beam produced with a Penning 
source was measured. Due to low count rates for the cold ion beams (only 10\% transmission), insufficient statistics could be gathered around the 
10-eV resonance. Therefore, the rate curves have been normalised to the 12-meV resonance. This resonance is at an energy low enough for sufficient 
statistics, however, presumably at an energy high enough for the normal- and para-\hhh\ rates to be the same, as observed in Ref. 
\cite{kreckel:2005a}. The hot \hhh\ rate curve, which does contain enough statistics, is normalised to the 10-eV resonance. Both normalisations are to 
the photocathode measurement of Ref. \cite{kreckel:2005a} as it bears a similar energy resolution. Figures \ref{fig:Trates} and \ref{fig:rates} show 
the reduced rate coefficients, $\alpha \cdot \sqrt{E_d}$, with $\alpha$ the energy-dependent rate coefficient that is proportional to $\sqrt{E_d}$. 
Any structures observable on the reduced rate are related to resonant behaviour. Figure \ref{fig:Trates} (upper frame) displays the measured rates of 
the hot and cold \hhh\ produced with normal-\htwo. Overall, the hot \hhh\ rate is higher than the cold \hhh\ rate, except for the resonance around 7 
and 270 meV. Although the temperature of the \hhh\ beam from the Penning source is on the order of $10^3$\,K (see Section \ref{sec:dyn}), resonant 
behaviour is still clearly visible. Figure \ref{fig:Trates} (lower frame) shows the theoretical predictions for 100\,K and 1000\,K as calculated by Ref. 
\cite{santos:2007}. Similar to the experiment, the hot \hhh\ rate is overall higher, however, at very low energies, the opposite behaviour is 
predicted. Several structures appear both in theory and in experiment. The intensities of the theoretical rates seem to be systematically lower than 
those measured, leading to a better agreement between our cold \hhh\ beam and the 1000\,K theoretical rate. This raises the question whether heating 
effects in the storage ring might have considerably warmed up the initially cold ion beam. The fragmentation dynamics presented in the next section, 
however, demonstrate that this is not the case.

Figure \ref{fig:rates} displays the reduced rates of the cold \hhh\ ion beams produced with normal- and para-\htwo\ (upper frame). The \hhh\ rate 
using para-\htwo\ exhibits slightly more structure below 5 meV collision energies. Otherwise, the rate coefficients are quite similar and show no sign 
of the predicted order of magnitude difference in DR rate between para-\hhh\ and ortho-\hhh\ \cite{santos:2007}. The two \hhh\ populations in the 
storage ring produced with normal- and para-\htwo\ are unknown and may very well differ less than was the case in the spectroscopy measurements, 
however, a DR rate that is an order of magnitude faster for para-\hhh\ than for ortho-\hhh\ should manifest itself even at small increases in the 
para-\hhh\ contribution. No large dissimilarities between the two source gases are observed that would point towards the predicted ten times more 
rapid dissociation of para-\hhh. Still, nearly identical \hhh($J,G$) populations in the storage ring for both source gases cannot be excluded.

Also shown in figure \ref{fig:rates} is the previous photocathode measurement for normal-\hhh\ (upper frame) and the previously measured \hhh\ rates 
for the normal- and para-\htwo\ source gases as measured with the thermal cathode (lower frame) \cite{kreckel:2005a}. The photocathode data clearly 
exhibit more pronounced resonances as the resolution of the liquid-nitrogen cooled photocathode (T$_{e,\perp}=1$ meV) \cite{orlov:2004} is much 
better than that of the thermal cathode (T$_{e,\perp}=4$ meV). A small resonance is revealed around 17 meV and below 5 meV structure is unfolded. The 
increased structure visible for the para-\htwo\ source gas raises the \hhh\ rate slightly, in accordance with the previous thermal-cathode experiment, 
i.e., assuming the normalisation on the 12-meV resonance is valid. Further analysis is still required. It is noted that the internal distributions in 
the ring may also differ between the present and previous DR experiments, both in \otp\ ratio for the \hhh\ produced with para-\htwo\ as in the 
achieved internal temperature that may give rise to a differing contribution from excited rotational levels irrespective of source gas.

\subsection{Dissociative Recombination Dynamics \label{sec:dyn}}

\begin{figure}[t]
\begin{minipage}{\miniwidth}
\includegraphics[angle=0,width=\textwidth,height=13pc]{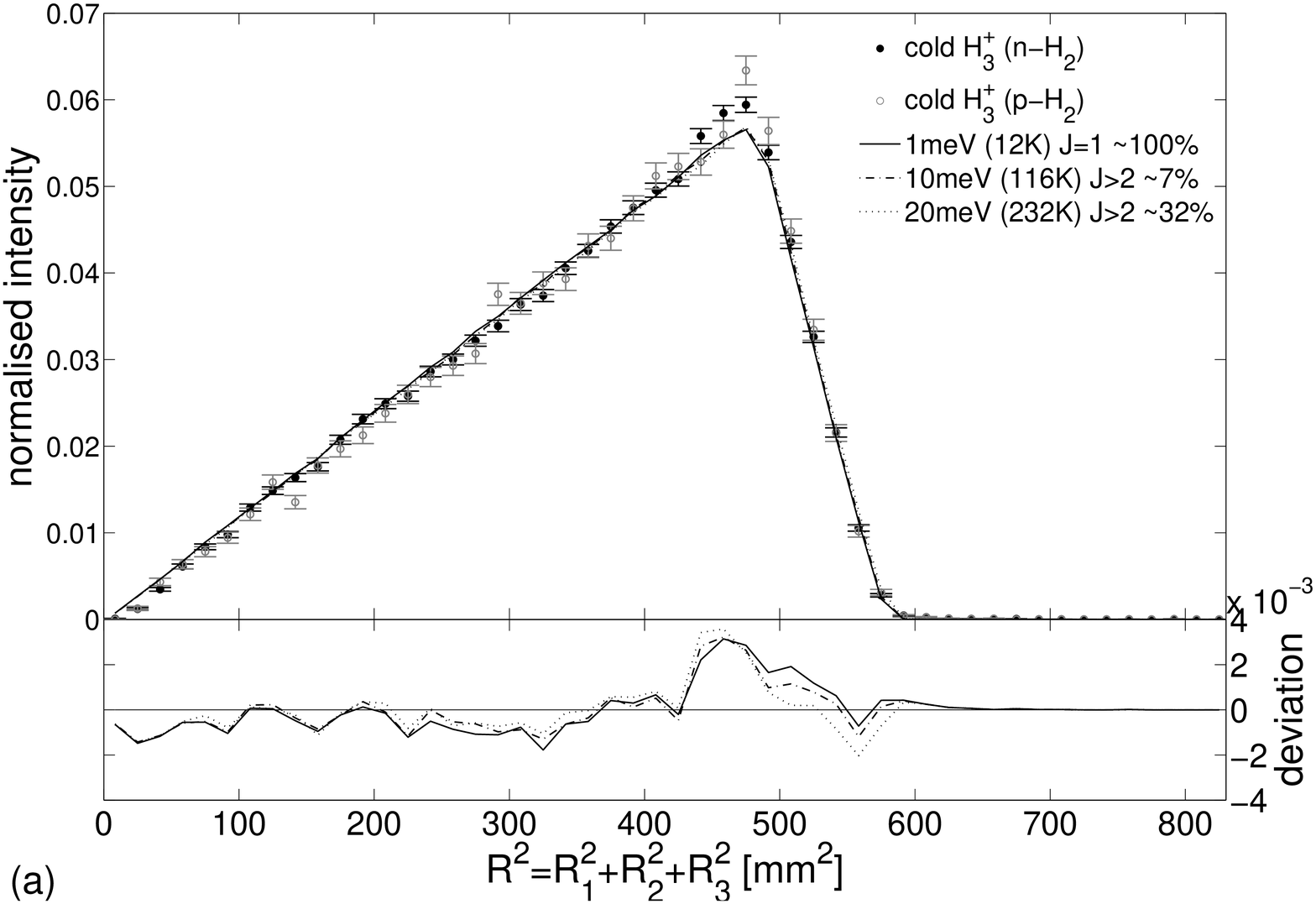}
\end{minipage}\hspace{1pc}%
\begin{minipage}{\miniwidth}
\includegraphics[angle=0,width=\textwidth,height=13pc]{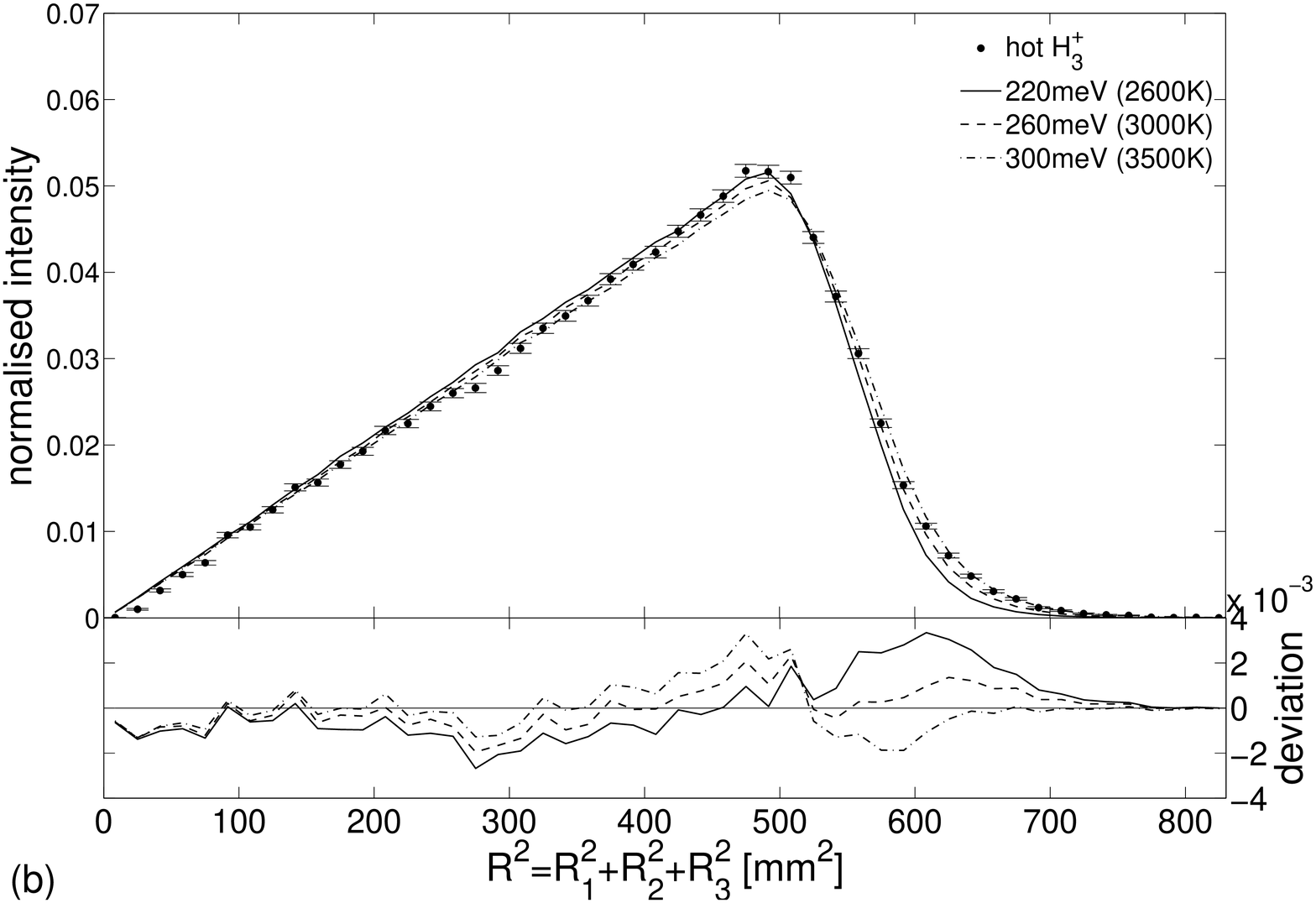}
\end{minipage}
\caption{\label{fig:2D} The $R^2$ distributions of the \hhh\ beams for (a) the cold \hhh\ produced with normal- and para-\htwo\ in the 22-pole trap 
and (b) the hot \hhh\ produced with normal-\htwo\ in the Penning source. All black curves are Monte-Carlo simulations of isotropic dissociations at 
the given temperatures, considering the allowed rotational levels and their probability in a Boltzmann distribution. All distributions are normalised 
to unit area. The deviation of the measured distributions of \hhh\ produced with normal-\htwo\ from the simulations are plotted in the respective 
lower frames.}
\end{figure}

The breakup dynamics of the DR of \hhh\ and their possible ($J,G$) state-dependences have been studied as well. Imaging data have been taken for both 
the 2-body and 3-body fragmentation channel. Here, only dissociation into three $H$ atoms is treated. Figure \ref{fig:2D}a) shows the projected 
distance distributions of our cold \hhh\ beams from the 22-pole trap as acquired by the $2D$ imaging \cite{strasser:2002a}, normalised to unit area. 
The distance plotted is $R^2=R_1^2 + R_2^2 + R_3^2$ with $R_i$ the distance from particle $i$ to the centre-of-mass in the $2D$ detection plane. 
Forward Monte-Carlo simulations for isotropic dissociations at three different rotational temperatures, normalised to unit area, are plotted to reveal 
a rough estimate of the temperature of the ion beam and anisotropy effects. To clarify differences, the deviation of the distribution 
for the normal-\htwo\ source gas from the simulations is displayed in the lower frame. The discrepancy at the peak of the distribution could indicate 
anisotropy and further simulations are needed to investigate the observed feature. Figure \ref{fig:2D}b) shows the $R^2$ distribution of the hot 
\hhh\ from the Penning source together with forward simulations at three high temperatures. Below, the difference of the measured data with the 
simulations is shown. Again, no large anisotropies can be observed.

The best temperature fit is determined from the least-squares minimum between the falling slope of the data and the simulation, with the simulation 
normalised to the maximum of the measured data (not displayed) in order to disregard the anisotropy effects. The hot \hhh\ distribution appears to 
have a temperature of around 3000\,K. For the cold \hhh, the least-squares deviation is smallest at the lowest temperature and increases with rising 
temperature. Although a significant contribution of the J=2 level cannot be excluded, the temperature is most certainly below room temperature, which 
demonstrates that the ion beams are indeed cold. The apparent difference between experiment and theory at low temperatures can therefore not be 
explained by considerable heating of the ion beam. 

\begin{figure}[t]
\begin{minipage}[b]{\miniwidth}
\includegraphics[angle=0,width=\textwidth]{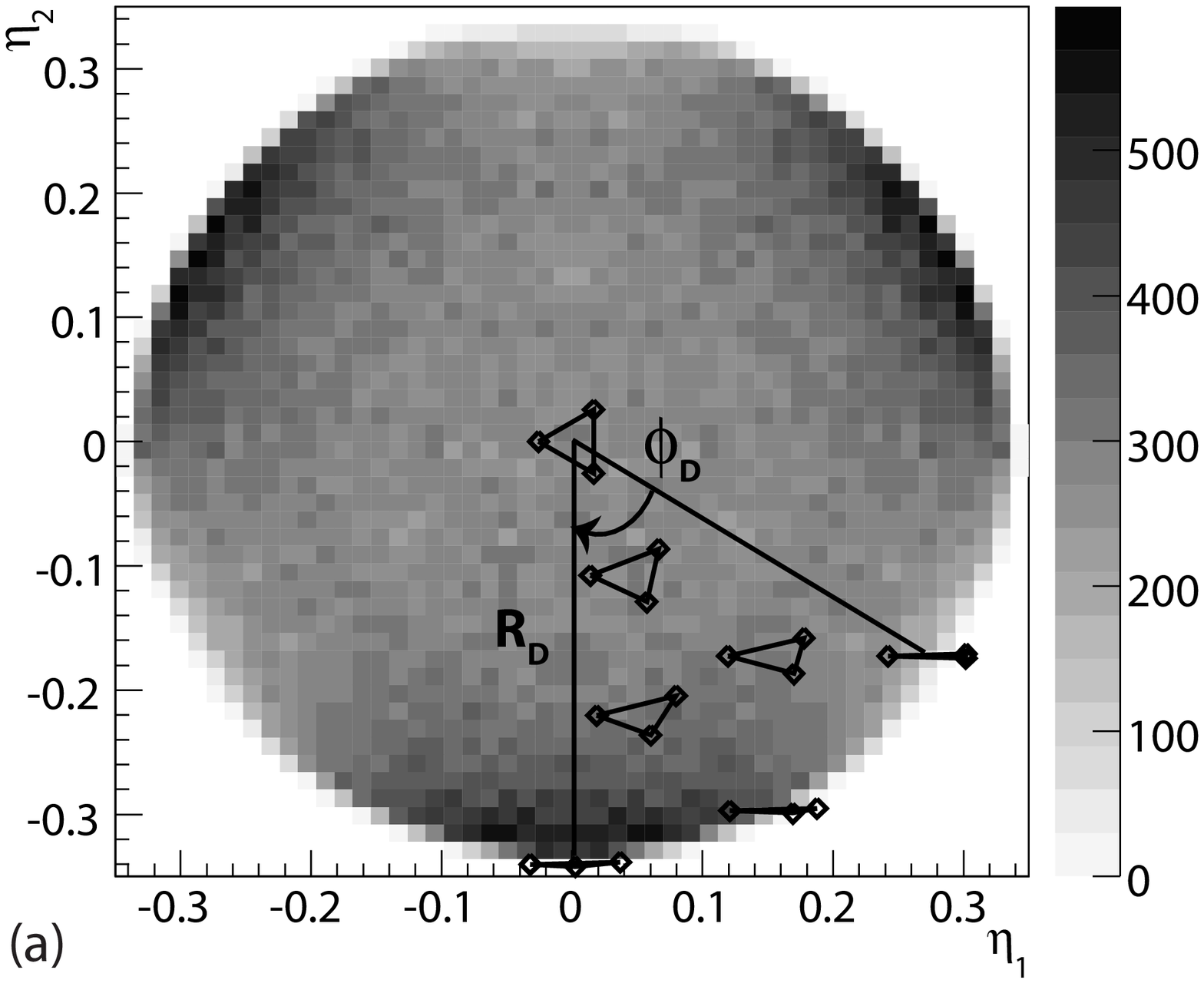}
\end{minipage}
\begin{minipage}[b]{\miniwidth}
\includegraphics[angle=0,width=\textwidth]{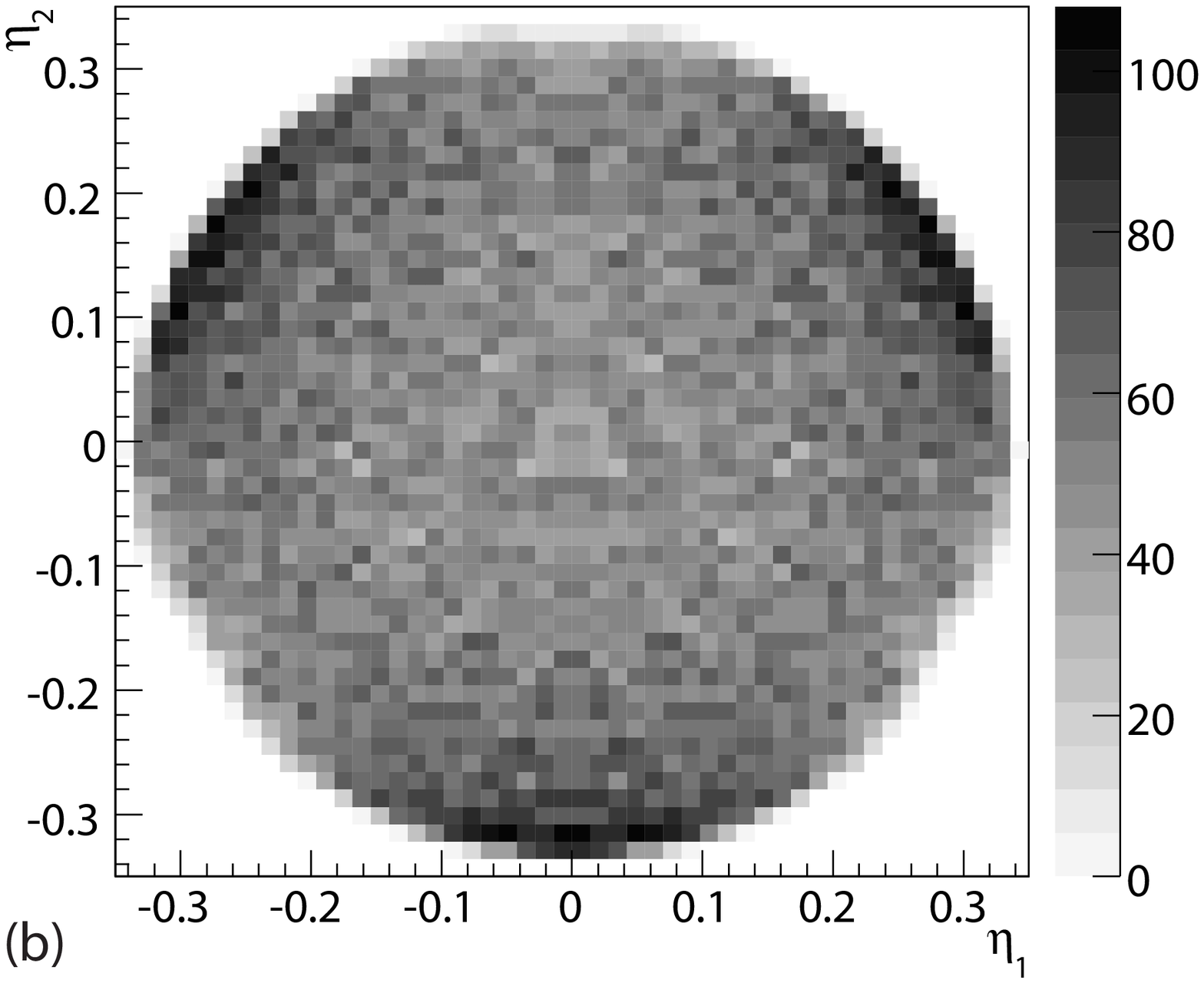}
\end{minipage}\hspace{1pc}%
\begin{minipage}[b]{\textwidth}
\caption{\label{fig:3D}The $3D$ imaging data of hot \hhh\ from the Penning source for a) all dissociations and b) the near-transversal dissociations, 
i.e., dissociations with $cos(\theta)$$>$0.8, where $\theta$ is the angle between the normal of the molecular plane and the beam axis.}
\end{minipage}
\end{figure}

Figure \ref{fig:3D}a) gives the Dalitz plot of the hot \hhh\ beam from the Penning source for 0-eV collision energy, for the first time acquired by 
the $3D$ imaging technique \cite{strasser:2000}. The geometries of the corresponding dissociations are displayed in the lower sextant. As 
proof-of-principle for the $3D$ imaging, the events dissociating near-perpendicular to the detector are displayed in figure \ref{fig:3D}b). 
Near-perpendicular events with $cos(\theta)$$>$0.8 are selected, where $\theta$ is the angle between the normal of the molecular plane and the beam 
axis. The tranversal $3D$ Dalitz plot is indeed similar to the total $3D$ Dalitz plot. The $3D$ imaging data are also consistent with the Dalitz plots 
obtained previously by Monte-Carlo reconstructions of $2D$ projected distance distributions \cite{strasser:2002a}.

\begin{figure}[h]
\includegraphics[angle=0,width=\textwidth]{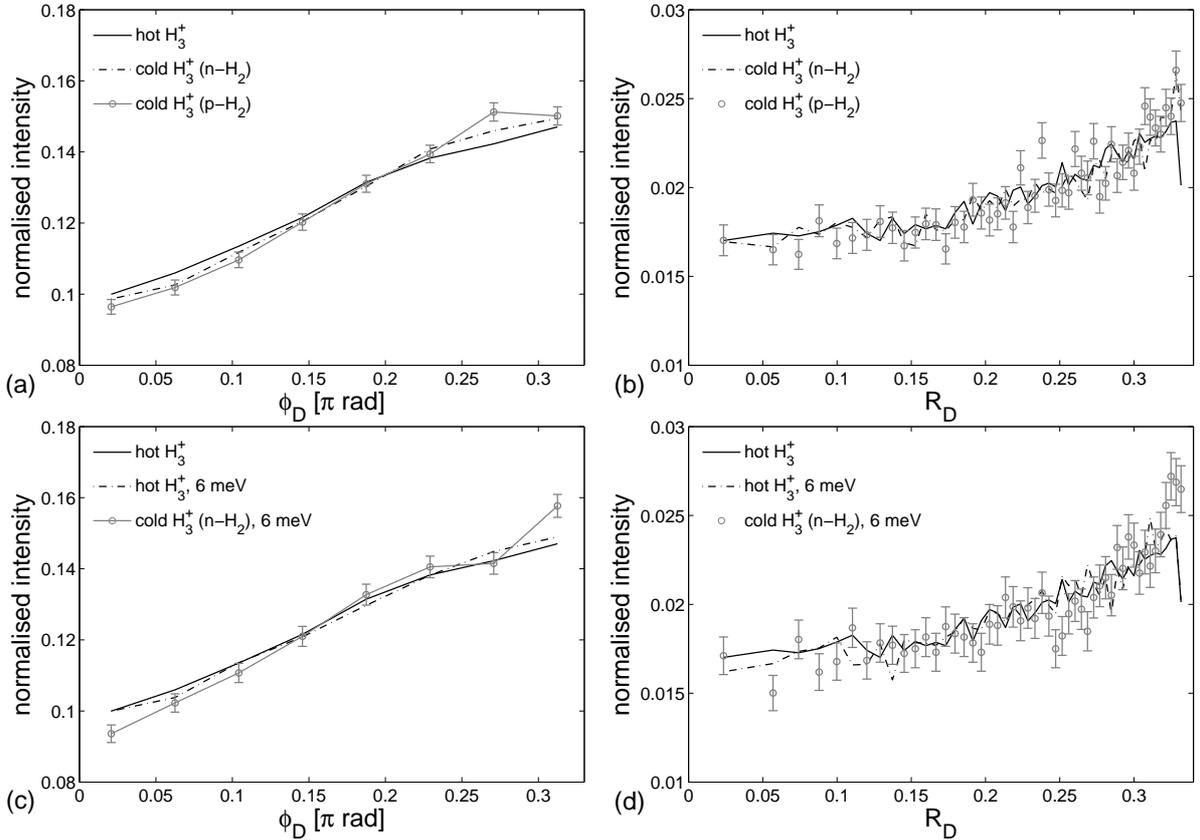}
\caption{\label{fig:3Dproj}The projected Dalitz polar coordinates at a,b) 0 eV and c,d) 6 meV for the given \hhh\ populations. For comparison, the hot 
\hhh\ data at 0 eV are displayed again with the 6 meV distributions. As illustrated in figure \ref{fig:3D}, R$_D$ ranges from the equilateral triangle 
to linear dissociation geometries and $\phi_D$ ranges from asymmetric to symmetric geometries. To improve clarity, only the errorbars of the 
distribution with the lowest statistics are shown.}
\end{figure}

To investigate the distributions for the different \hhh\ populations in more detail, the Dalitz coordinates have been transformed to polar coordinates 
(as illustrated in figure \ref{fig:3D}a) to be able to integrate over the angle and the radius. The integrated radius describes the amount of 
(non-)linearity and the integrated angle the deformation from the isosceles/equilateral triangle. Figures \ref{fig:3Dproj}a) and b) display the 
integrated-coordinate distributions at 0 eV. A uniform background, i.e., isotropic distributed geometries, can be observed superimposed with 
$\sim$12\% anisotropic dissociations, which exhibit a preference towards linear breakup, independent of the \hhh\ population. As discernable in the 
Dalitz plot, the near-linear dissociations peak around the symmetric linear breakup and drop to (even slightly below) the level of the isotropic 
background for the asymmetric linear geometries. The integrated-coordinate  distributions show no significant internal-state dependence. Figures 
\ref{fig:3Dproj}c) and d) display the distributions for the hot and the cold \hhh\ at 6 meV together with the 0-eV distributions of the hot \hhh\ for 
comparison. The breakup of the cold \hhh\ produced with para-\htwo\ at 6 meV was not measured. The fragmentation dynamics of the hot \hhh\ at 6 meV 
resemble those at 0 eV. The cold \hhh\ breakup exhibits a small difference, peaking slightly more around the symmetric linear dissociations. No large 
energy dependence is observed.

\section{Conclusions \label{sec:conc}}
The rovibrational spectroscopy shows that the use of para-\htwo\ as parent gas can indeed enrich the para contribution of the \hhh\ population. In the 
storage-ring experiment, where the para enrichment cannot be confirmed, no increase in the DR rate has been found when using para-\htwo\ as source gas 
that may support the prediction of a 10 times higher para-\hhh\ rate than ortho-\hhh\ rate. The trend of increasing overall rate with increasing 
temperature as predicted by theory is, however, confirmed by the present data. Furthermore, the \hhh\ ions prepared in the 22-pole trap and 
transferred to the TSR are really rovibrationally cold, featuring a temperature that is distinctly lower than 300\,K. As such, heating effects cannot 
account for the discrepancy in intensity between the theoretical and experimental rate-coefficient curves. The 3-body fragmentation dynamics of the DR 
of \hhh\ show no large dependences on the internal-state distribution of the initial \hhh\ nor on electron energy. A fraction of 12\% of the \hhh\ 
ions are found to dissociate in symmetric near-linear geometries, while 88\% have a uniform distribution of geometries.

\section*{Acknowledgments}
HB acknowledges partial support from the German Israeli Foundation for Scientific Research and Development (G.I.F.) under Grant I-900-231.7/2005 and 
by the European Project ITS LEIF (HRPI-CT-2005-026015). We are grateful to the TSR accelerator group for their support during the experiments.

\section*{References}
\bibliography{DRproceedings}

\end{document}